\documentclass[showpacs,english,aps,pra,twocolumn,footinbib,superscriptaddress,floatfix]{revtex4}

\usepackage[dvips]{graphicx}
\usepackage{babel,amsmath,amssymb,float}
\usepackage{pstricks}
\usepackage{color}

\newcommand{\ket}[1]{|#1\rangle}

\newcommand{\op}[1]{\hat{\mathrm{#1}}}
\renewcommand{\check}{\op{Z}}
\newcommand{\prob}{\mathcal{P}}
\newcommand{\chain}{E}

\newcommand{\hamiltonian}{\mathcal{H}}

\newcommand{\sset}[1]{ \{#1\} }

\definecolor{yblue}{rgb}{0.06, 0.3, 0.57}
\usepackage[pdftex]{hyperref}
\hypersetup{colorlinks=true,linkcolor=yblue,citecolor=yblue,urlcolor=yblue}

\begin{document}

\title{Error tolerance of topological codes with independent bit-flip
and measurement errors}

\author{Ruben S.~Andrist}
\affiliation{Santa Fe Institute, 1399 Hyde Park Road, Santa Fe, New Mexico
87501, USA}

\author{Helmut G.~Katzgraber}
\affiliation{Department of Physics and Astronomy, Texas A\&M University,
College Station, Texas 77843-4242, USA}
\affiliation{Santa Fe Institute, 1399 Hyde Park Road, Santa Fe, New Mexico
87501, USA}

\author{H.~Bombin}
\affiliation{Yukawa Institute for Theoretical Physics, Kyoto University,
Sakyo, Kyoto 606-8502, Japan}

\author{M.~A.~Martin-Delgado}
\affiliation{Departamento de F{\'i}sica Te{\'o}rica I, Universidad
Complutense, 28040 Madrid, Spain}

\begin{abstract}

Topological quantum error correction codes are currently among the most
promising candidates for efficiently dealing with the decoherence
effects inherently present in quantum devices.  Numerically, their
theoretical error threshold can be calculated by mapping the underlying
quantum problem to a related classical statistical-mechanical spin
system with quenched disorder.  Here, we present results for the general
fault-tolerant regime, where we consider both qubit and measurement
errors. However, unlike in previous studies, here we vary the strength
of the different error sources independently. Our results highlight
peculiar differences between toric and color codes. This study
complements previous results published in New J.~Phys.~{\bf 13}, 083006
(2011).

\end{abstract}


\maketitle

\label{sec:intro}

The quest for building a reliable quantum computer involves multiple
fields of research, such as several branches of computer science,
theoretical and experimental physics, mathematics, and engineering
\cite{nielsen:00,galindo:02}.  Most notably, disordered spin systems and
lattice gauge theories \cite{dennis:02,ohno:04,wang:03,andrist:10} in
statistical physics have played a pivotal role in the understanding of
the theoretical error tolerance of topological quantum computing models
in quantum information theory
\cite{katzgraber:09c,ohzeki:09a,dennis:02,raussendorf:07,fowler:11,landahl:11}.
The main driver for this fruitful synergy across disciplines was the
discovery that methods from statistical physics allow for the numerical
study of quantum error correction codes
\cite{shor:95,steane:96,calderbank:96,steane:96a,gottesman:96,bennett:96a,terhal:15}.
More specifically, error fluctuations in topologically protected codes
map \cite{dennis:02} directly  onto classical spin models with tunable
disorder. The level of noise in the quantum code then corresponds to the
amount of quenched disorder in a classical spin system. In practice,
this means that by carefully analyzing the critical behavior of the
classical system, we can learn how resilient a topological code is to a
particular source of errors.

The very same approach can also be used to investigate fault-tolerant
error correction \cite{shor:96,knill:96,aharonov:97}, which takes the
possibility of faulty measurements during the error-correction procedure
into account. This is particularly exciting because topological codes
\cite{kitaev:03} allow for fault-tolerant quantum computation without
resorting to code concatenation
\cite{dennis:02,bombin:06,bombin:07b,bombin:07,nayak:08}. Instead, the
new resource is the nontrivial topology of the lattice on which the
physical qubits are arranged. The topological quantum code, in turn, is
defined by the pattern of the arrangement and the way in which check
operators act on these qubits. The key property of these check operators
(also known as stabilizers), is that their support is local on the
qubits forming the lattice. This locality property is absent in
concatenated codes and is beneficial for experimental realizations.
Moreover, as long as the external errors act also locally on the code,
it is possible to protect the encoded quantum information because the
encoded logical qubits are entangled states that spread out globally
across the whole system. While implementing such systems might seem to
be an insurmountable effort at this time, recently, a complete
error-correction code for arbitrary errors using a minimal topological
color code has been realized experimentally in a trapped-ion platform
\cite{nigg:14a} that paves the way towards the experimental realization
of topological codes, such as the Kitaev code or the color codes in
two-dimensional setups
\cite{niedermayr:14,barends:14,corcoles:15,kueng:15,mueller:16}.
As such, gaining a deeper understanding of the interplay
between different error sources is of current importance.

Remarkably, assuming that both bit-flip errors and measurement errors
occur at the same average rate, previous numerical results
\cite{andrist:10} suggest that topological color codes \cite{bombin:06}
exhibit an improved error tolerance over the toric code
\cite{kitaev:03}.  While this is potentially only true in the ideal
scenario where all physical operations are noise free, it does serve as
a guide when comparing the performance of both models on an equal
footing.  Here, we further investigate this observation by extending the
numerical results to qubit and measurement errors of different average
strength.

The paper is organized as follows. Section \ref{sec:tolerance} provides
a brief introduction to the toric code and topological color codes in
the fault-tolerant regime. Section \ref{sec:mapping} summarizes the
mappings to classical lattice gauge theories as derived in
Ref.~\cite{andrist:10} for color codes and Ref.~\cite{ohno:04} for the
toric code. In Sec.~\ref{sec:numerical} we explain the numerical tools
used for our extended analysis, followed by the results in
Sec.~\ref{sec:results}, as well as concluding remarks.

\section{Topological stabilizer codes}
\label{sec:tolerance}

A stabilizer code ${\cal C}$ of length $n$ is a subspace of the Hilbert
space of a set of $n$ physical qubits \cite{gottesman:96}. The code is
defined by means of the stabilizer group ${\cal S }\subset {\cal P}_n$
of Pauli operators, which are tensor products of Pauli matrices of
length $n$:
\begin{equation}
	{\cal P}_n := \langle 1, X_1,Z_1, \ldots, X_n, Z_n \rangle.
\end{equation}
The stabilizer group leaves invariant the quantum states belonging to
the code:
\begin{equation}
	\ket \psi \in \mathcal C \qquad \iff \qquad \forall\, O\in \mathcal
	S\quad O\ket \psi = \ket \psi.
\end{equation}
The Pauli operator $-1$ is excluded from ${\cal S}$. To fully
characterize the code it is sufficient to define the generators of
${\cal S}$:
\begin{equation}
	{\cal G} := \langle 1, g_1, \ldots, g_k \rangle.
\end{equation}
The normalizer $\mathcal N(S)$ of ${\cal S}$ plays a fundamental role in
error correction. It is defined by the operators $O$ satisfying:
\begin{equation}
	O \in \mathcal N(S) \qquad \iff \qquad O \mathcal S = \mathcal S O,
\end{equation}
which implies that the code space $\mathcal C$ is left invariant by
$\mathcal N(S)$. When the operators of the normalizer do not belong to
the stabilizer itself, then they act in a non-trivial way on the encoded
states.

Active error correction is necessary to protect the error-prone logical
state: we need to measure a set of generators of $\mathcal S$. The
result of these measurements is called the syndrome---the signature of
which error has occurred. Errors can be corrected as long as the
syndromes allow us to discriminate among possible errors. As correctable
errors always form a vector space, it is enough to consider Pauli
operators, which form a basis. A Pauli error $e$ is said to be
undetectable if it belongs to the set $\mathcal N(S) - \mathcal S$. In
this case, the syndrome provides no information:
\begin{equation}
	\forall\, s \in \mathcal S \qquad s\ e \ket{\psi} = e \ s' \ket{\psi}
	= \ket{\psi}.
\end{equation}
A set of Pauli errors $E$ is said to be correctable if, and only if,
\begin{equation}
	E^{\dagger} E \cap \mathcal N(S) \in \mathcal S.
\end{equation}
Topological stabilizer codes are peculiar instances of stabilizer codes
employing a regular arrangement of qubits on a topologically non-trivial
surface \cite{bombin:07d} and local stabilizer operators. Because both
codes are Calderbank-Shor-Steane codes \cite{calderbank:96,steane:96a},
bit-flip and phase errors can be corrected independently and
analogously. Here we focus only on bit-flip errors occurring at a rate
$p$.

\paragraph*{Toric code.} The physical qubits are arranged on the
edges of a two-dimensional lattice, with stabilizers at each vertex
being the tensor product of $\check$ operators for adjacent qubits
\cite{kitaev:03}.  Thus flipping qubit $Q_\ell$ changes the sign of the
measured eigenvalue for the check operators at either end of the edge
$\ell$.  The first example of such a topological code was the Kitaev
toric code defined with a square-lattice arrangement. In this case, each
stabilizer operator $\check^{\otimes4}$ is the tensor product of exactly
four $\check$ operators.

\paragraph*{Topological color codes.} Initially conceived to expand
the computational capabilities of topological codes by increasing the
set of topological gates that can be applied
\cite{bombin:06,bombin:07b}, here we consider a hexagonal arrangement of
the physical qubits, with stabilizers $\check^{\otimes6}$ on each
plaquette acting on the adjacent qubits.

\paragraph*{Error correction.} For all check operators, encoded
states satisfy $\check \ket{\psi} = \ket{\psi}$. Such states exist
because the group generated by check operators, called the stabilizer
group, does not contain $-\mathbf{1}$, so that in particular check
operators commute with each other. The dimension of the encoded subspace
depends only on the topology of the surface where the code lives. For
example, a regular lattice with periodic boundary conditions has the
topology of a torus and encodes two logical qubits \cite{bombin:06}.

\paragraph*{Fault-tolerant regime.} With measurements being faulty at
a rate $q$, new errors are introduced involuntarily during the
error-correction procedure.  To detect local inconsistencies with the
code, check operators need to be measured repeatedly over time and error
correction amounts to guessing the correct error history $\chain$ among
those that are compatible with the recorded measurement outcomes. Such
an error history is typically comprised of some combination of
bit-flip and measurement errors. Indeed, many error histories have an
equivalent effect, and thus the ideal strategy is to compute which
equivalence class $\bar \chain$ happened with the highest probability
$P(\bar \chain)$. Therefore, error correction is highly successful when
for typical errors there is a class that dominates over the others.

\section{Mapping to lattice gauge theories}
\label{sec:mapping}

\begin{figure}
\includegraphics[width=\columnwidth]{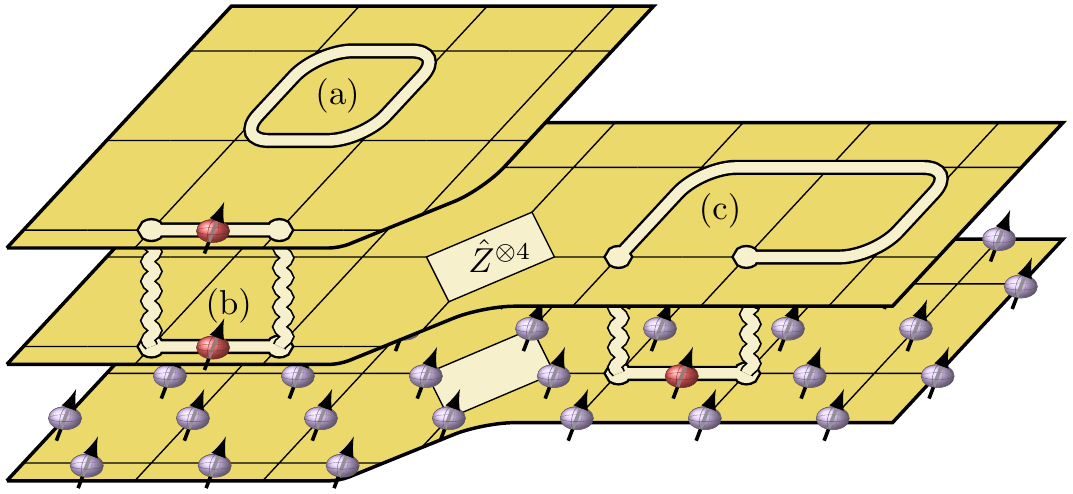}
\vspace{.3mm}
\caption{(Color online)
Stacked layers representing the mapped model for the fault-tolerant
toric code. Qubits reside on the edges and stabilizer operators
$\check^{\otimes4}$ act on the qubits surrounding each vertex.  (a)
Horizontal loops correspond to the usual local equivalence of the toric
code: flipping the four qubits around a plaquette leaves the error
syndrome invariant.  (b) The second type of local equivalence involves
measurement errors which are represented by vertical links connecting
stabilizer operators.  (c) The resulting model consists of spatial and
time-like links forming a three-dimensional cubic lattice.
}
\label{fig:toric:lattice}
\end{figure}

For both types of topological stabilizer codes introduced in the
previous section, the mapping of the setup to a classical spin system
produces a lattice gauge theory with
disorder \cite{dennis:02,andrist:10}. For a side-by-side comparison, it
is instructive to describe both in terms of their local equivalences.
The results of the respective mappings (with some minor adjustments to
match the notations) can be summarized as follows.

\begin{figure}[b!]
\includegraphics[width=\columnwidth]{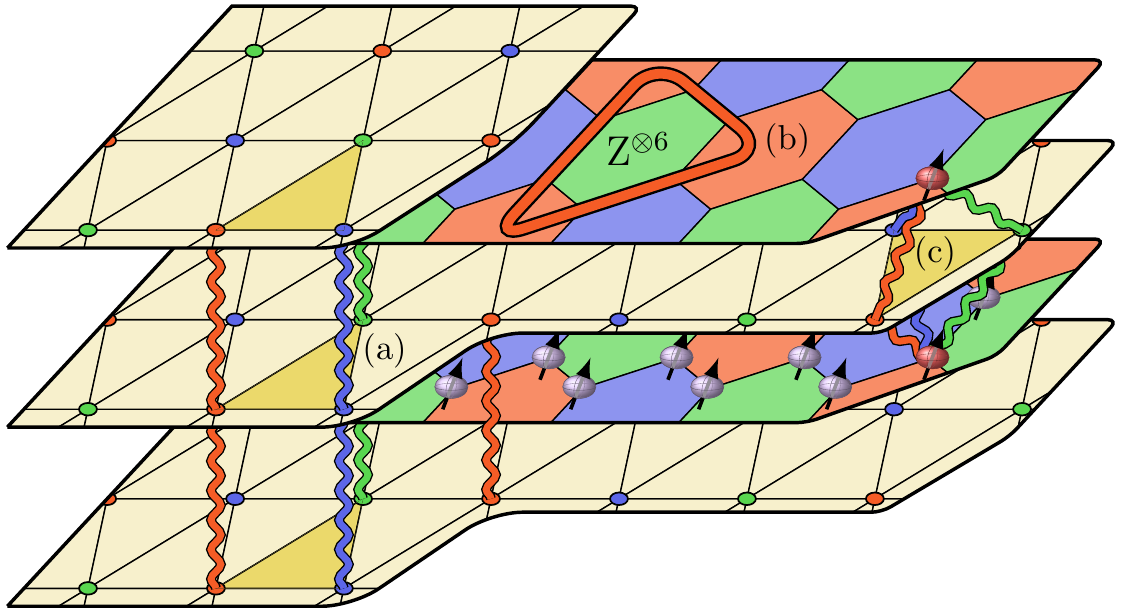}
\caption{(Color online)
Resulting lattice gauge theory for topological color codes on a
hexagonal lattice. (a) The horizontal layers are alternating triangular
and hexagonal lattices, with time corresponding to the vertical axis.
Qubits reside on the vertices of the trivalent (hexagonal) lattice.  (b)
Colored loops in the hexagonal planes correspond to bit-flip error
chains.  (c) Vertical loops involve measurement errors between two time
slices.  This scenario is analogous to the case of two unnoticed
consecutive errors for the toric code.
}
\label{fig:color:lattice}
\end{figure}

\paragraph*{Toric codes (${\mathbb Z}_2$ lattice gauge theory).} The
toric code for bit-flip errors occurs on the edges of a square lattice
with stabilizer operators at each vertex acting on the four adjacent
qubits (see Fig.~\ref{fig:toric:lattice}). Fault tolerance is added by
stacking multiple of these lattices and connecting the vertices
vertically to a three-dimensional cubic grid. We can then interpret the
vertical axis as time with each vertical edge representing a measurement
of the stabilizer operator it connects.

In addition to the regular local equivalence of flipping four qubits
around a horizontal plaquette, this stacked model also has a second type
that consists of flipping the same qubit in adjacent layers along with
the two measurements connecting them. This equivalence is a vertical
plaquette in the three-dimensional lattice and represents the scenario
of two consecutive bit-flip errors which go unnoticed because of two
concurrent measurement errors.

The probability of an error arbitrary history $\chain$, consisting of
$h$ bit-flip errors and $v$ faulty measurements, can be written as
\begin{align}
	\prob(\chain)&=(1-p)^{H-h}\,p^h (1-q)^{V-v}q^v\notag\\
	&\propto \left (\frac p{1-p}\right)^h \left (\frac q{1-q}\right)^v\,,
    \label{eq:toric:prob}
\end{align}
where $p$ is the bit-flip rate and $q$ the measurement error rate,
while both $H$ (total number of qubits) and $V$ (total number of
measurements) are constants of the cubic lattice.

A specific error history $\chain$ can be represented by a set of
variables $\tau_\ell\in\sset{\pm1}$, each indicating whether the qubit
or measurement corresponding to edge $\ell$ is faulty. Furthermore, we
can enumerate all histories in the error class of $\chain$ (i.e., those
that differ only by local equivalences) by attaching a binary variable
$\sigma_{h,v}\in{\pm1}$ to each equivalence.  To numerically sample from
these, one then constructs a classical Hamiltonian which has Boltzmann
weights proportional to Eq.~\eqref{eq:toric:prob}:
\begin{equation}
	\hamiltonian_E =
		-J \sum_{j\in \ell_Q} \tau_{j} 
			\sigma_h^{\otimes 2} \sigma_v^{\otimes 2}
		-K \sum_{k\in \ell_M} \tau_{k} \sigma_v^{\otimes 4}\,,
	\label{eq:toric:hamiltonian}
\end{equation}
Note that the first sum (which iterates over all qubits $Q$, i.e.,
horizontal links) essentially counts the number of flipped qubits. By
definition, a qubit is flipped if the product of $\tau_j$ and all the
equivalences it is affected by (two horizontal and two vertical ones) is
negative. Similarly, the second sum iterates over all measurements $M$
and adds up the number of faulty ones.  Therefore, we can see that the
correct Boltzmann weights are produced with
\begin{align}
    e^{-2\beta J}&= p/(1-p)\,,&
    e^{-2\beta K}&= q/(1-q)\,,
    \label{eq:nishimori}
\end{align}
which is called the Nishimori condition \cite{nishimori:81}. The
Hamiltonian in Eq.~\eqref{eq:toric:hamiltonian} is equivalent to the one
given by Dennis {\em et~al.} \cite{dennis:02}, however, with separated
terms for qubit and measurement errors.

{\em Color codes (tricolored lattice gauge theory).} For topological
color codes, consider a three-dimensional lattice consisting of stacked
triangular and hexagonal layers, with qubits residing on intermediate
hexagonal layers. There is a stabilizer operator $\check^{\otimes6}$ for
each of the hexagonal tiles, acting on the six qubits surrounding the
plaquette.

As for the toric code, there are again two distinct types of elementary
equivalences. The first is a horizontal loop consisting of the six
qubits around a plaquette, while the second consists of adjacent qubits
in two layers, connected by three measurement errors (see
Fig.~\ref{fig:color:lattice}). This represents again the scenario of two
subsequent qubit flips on the same qubit, which remain unnoticed because
of three concurrent measurement errors. The resulting Hamiltonian takes
the form
\begin{equation}
	\hamiltonian_E =
		-J \sum_{j\in Q} \tau_{j} \sigma_{h}^{\otimes 3}
			\sigma_{v}^{\otimes 2}
		-K \sum_{k\in M} \tau_{k} \sigma_{v}^{\otimes 6}\,,
	\label{eq:color:hamiltonian}
\end{equation}
with identical requirements for the constants $J$ and $K$. This
corresponds to the Hamiltonian calculated by Andrist~{\em et
al.} (See Refs.~\cite{andrist:10,andrist:12} for details).

\section{Numerical Methods}
\label{sec:numerical}

Based on the Hamiltonians in Eq.~\eqref{eq:toric:hamiltonian} for the
toric code and Eq.~\eqref{eq:color:hamiltonian} for color codes, the
error threshold for a particular code is given by the largest error
rates for which the model remains in an ordered state at the temperature
$T$ specified by the Nishimori condition. Dennis {\em et
al.}~\cite{dennis:02} have demonstrated that this property is found at
the multicritical point of the $p$--$T$ ($p$ the bit-flip error
rate) phase diagram where the Nishimori line intersects the phase
boundary. For independent qubit and measurement error rates, the
Nishimori condition translates to a {\em Nishimori sheet} in the
three-dimensional parameter space spanned by $p$, $q$ and the model's
temperature $T=1/\beta$. Note that the purpose of this ``virtual''
temperature is merely to achieve the desired Boltzmann statistics via
Eq.~\eqref{eq:nishimori}, while any physical temperature effects in the
quantum device are implicitly captured by the error rates $p$ and $q$.

\begin{table}[!tb]
\caption{
Simulation parameters: $L$ is the layer size, $M$ is the number of
layers, $N_{\rm sa}$ is the number of disorder samples, $t_{\rm eq} =
2^{b}$ is the number of equilibration sweeps, $T_{\rm min}$ ($T_{\rm
max}$) is the lowest (highest) temperature, and $N_{\rm T}$ is the number
of temperatures used for a given error rate $p$. The corresponding
values of $q$ are given by the simulation paths chosen, namely
$q = 2p$, $q = p$, and $q = p/2$.
\label{tab:simparams}
}
\begin{tabular*}{\columnwidth}{@{\extracolsep{\fill}} l r r r r r r r}
\hline
\hline
$p$ & $L$ & $M$ & $N_{\rm sa}$ & $b$ & $T_{\rm min}$ & $T_{\rm max}$ &$N_{\rm
T}$ \\
\hline
$0.00$ & $6$, $9$ & $6$, $8$ & $1600$  &  $15$ & $1.20$ & $2.00$ & $64$\\
$0.00$ & $12$     & $12$     & $800$   &  $15$ & $1.20$ & $2.00$ & $64$\\
$0.02$ & $6$, $9$ & $6$, $8$ & $1600$  &  $16$ & $0.90$ & $1.80$ & $52$\\
$0.02$ & $12$     & $12$     & $800$   &  $17$ & $0.90$ & $1.80$ & $52$\\
$0.03$--$0.039$ & $6$, $9$ & $6$, $8$& $1600$ & $17$ & $0.70$ & $1.40$ & $52$\\
$0.03$--$0.039$ & $12$     & $12$    & $800$  & $19$ & $0.70$ & $1.40$ & $52$\\
$0.04$--$0.060$ & $6$, $9$ & $6$, $8$& $1600$ & $18$ & $0.50$ & $1.20$ & $52$\\
$0.04$--$0.060$ & $12$     & $12$    & $800$  & $20$ & $0.50$ & $1.20$ & $52$\\
\hline
\hline
\end{tabular*}
\end{table}

We use large-scale Monte Carlo simulations to analyze the phase diagram
along different projections, namely $p=2q$ and $p=q/2$. In both cases we
expect to find the system in an ordered Higgs phase for weak disorder
and low temperatures $T$. This indicates that error histories observed
at these error rates typically exhibit only small fluctuations. Once the
phase boundary is crossed, the system enters the disordered confinement
phase, indicating that the topologically encoded information is
vulnerable to failures.

\begin{figure}
\includegraphics[width=\columnwidth]{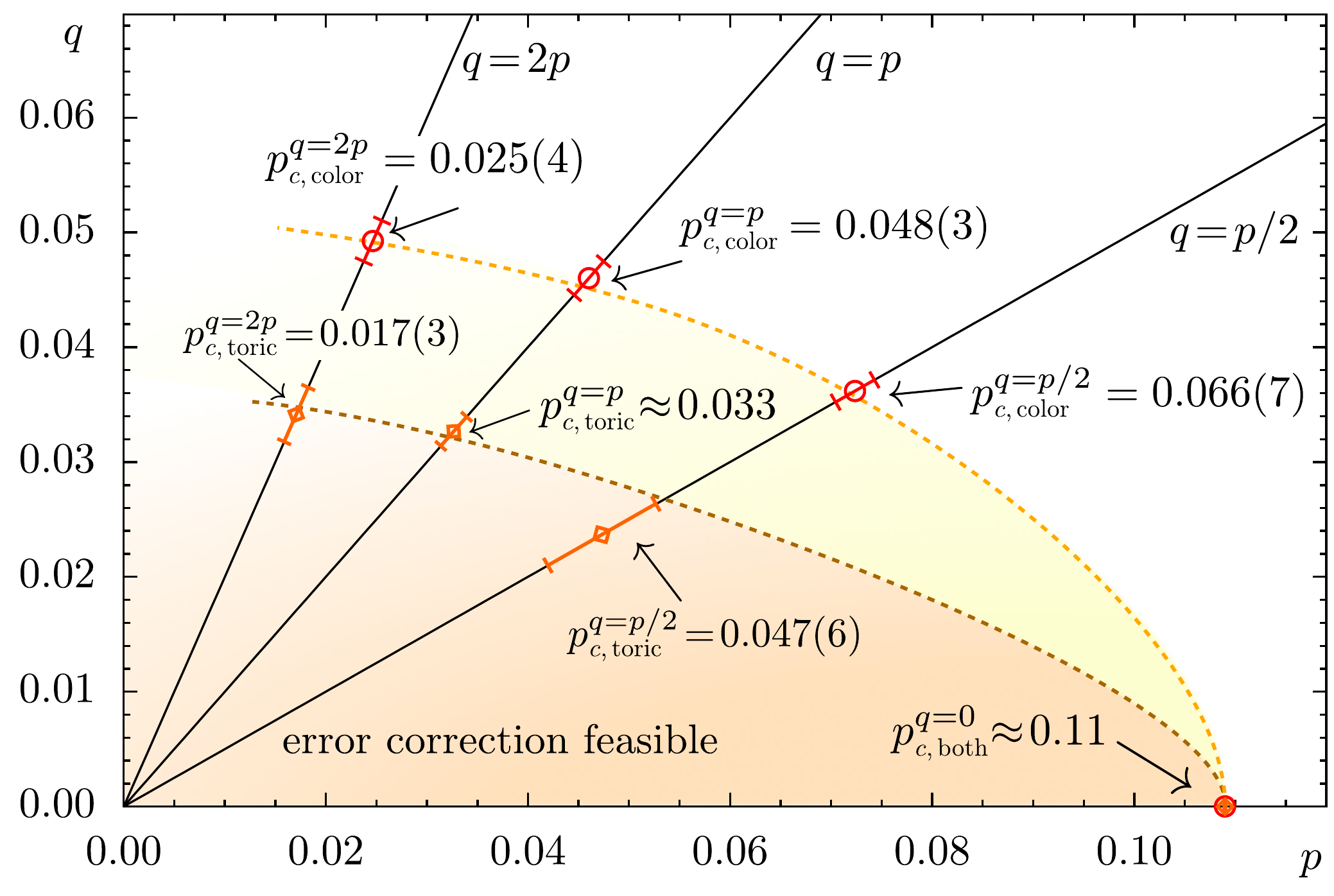}
\caption{(Color online)
Summary of the numerically calculated error thresholds in context of the
previous results. The plot indicates the phase boundaries of the ordered
phase for both types of codes, projected onto the Nishimori surface
where the mapping to the quantum setup is valid. The estimates indicate
that the difference in resilience remains even for non-matching error
rates. Simulations for $p \to 0$ are difficult and we have no estimates
for this regime.
}
\label{fig:pq_plane}
\end{figure}

The crossing point is determined as follows. For increasingly larger
error rates $p$ and $q$, we use the peak position in the measured
system's susceptibility as done in Ref.~\cite{ohno:04}, as well as the
skewness of the Wilson look distribution \cite{andrist:10} to locate the
phase transition temperature $T_c(p,q)$. As long as this transition
occurs at a higher temperature than the one specified by the Nishimori
condition, we know that the system still exhibits an ordered state.
Because the error rates $p$ and $q$ are merely parameters to generate
the quenched random interactions, these calculations need to be repeated
for many independent disorder realizations to obtain the desired
statistical-mechanical average. This and the fact that disordered
lattice gauge theories are inherently hard to simulate necessitates
considerable numerical efforts for every single point generated in the
phase diagram in Fig.~\ref{fig:pq_plane}.  To mitigate this challenge,
we use the parallel tempering Monte Carlo technique \cite{hukushima:96},
with the detailed simulation parameters listed in
Table~\ref{tab:simparams}. Equilibration for each sample is tested by a
logarithmic binning of the data. Once the last three bins agree within
statistical error bars, the system is deemed to be in thermal
equilibrium.

\section{Results}
\label{sec:results}

Our results are summarized in Fig.~\ref{fig:pq_plane}, which also
includes estimates of critical points computed in previous publications
\cite{dennis:02,ohno:04,katzgraber:09c,andrist:10}.  In principle, the
parametric space of the models constructed in Sec.~\ref{sec:mapping} is
three dimensional, spanned by $p$, $q$ and $T$.  However, since the
Nishimori surface represents the locus of points where the mapping from
the quantum setup is valid, we can render the results in a $p$--$q$
plot by projecting onto the Nishimori sheet along the temperature axis.
The colored areas then represent the portion of phase space for which
the Nishimori sheet is within the ordered phase of the model.

When measurement errors are not taken into account, it was shown in
Refs.~\cite{dennis:02,katzgraber:09c} that both codes have equal
thresholds of $p_c\approx 0.109$. This is indicated in
Fig.~\ref{fig:pq_plane} by both phases intercepting the horizontal axis
at the same point. Ohno~{\em et~al.}~\cite{ohno:04} have estimated the
threshold for the toric code at equal error rates to be
$p_c^{q=p}\approx 0.03$. Remarkably, the error resilience of topological
color codes under the same circumstances was found \cite{andrist:10} to
be substantially higher at $p_c^{q=p}\approx 0.048(3)$. To further the
understanding of this curious difference, we complement these previous
results with estimates for non-matching error rates at $q=2p$ and
$q=p/2$.  In both cases, there is still a notable difference between the
two types of codes. It thus seems that color codes are more resilient to
noise and measurement errors than toric codes across the whole Nishimori
sheet. The lines in Fig.~\ref{fig:pq_plane} are meant as guides to the
eye. Simulating more points in the phase diagram is extremely
difficult.  However, we feel that the results are robust within error
bars.

\section{Conclusions}
\label{sec:conclusions}

Our numerical results indicate that the difference in error resilience
between the toric code and topological color codes persists when
bit-flip and measurement errors occur at different rates. Only in the
limit of a vanishing measurement error rate the two lines in
Fig.~\ref{fig:pq_plane} converge to a common point, i.e., both toric and
color codes have the same error threshold to bit-flip errors. Exploring
the regime where measurement errors are far more common than bit-flip
errors is extremely difficult numerically because of the anisotropy of
the resulting lattice gauge theory. However, for a large portion of the
phase diagram in the $p$--$q$ plane both topological schemes show
different error tolerance. Gaining a complete understanding of the
underlying cause for the differences between the two types of
topological error-correction codes in the fault-tolerant regime will
require new analysis approaches by improving the error model with more
realistic features like taking into account the unavoidable noise
introduced by real physical operations during the correction protocol.
Whether the differences between both codes vanish under external noise
remains an open problem, and this may also require more detailed studies
of lattice gauge theories with quenched bond disorder. This, however, is
a numerically and analytically extremely challenging problem.

\section*{Acknowledgments} 

The authors thank ETH Zurich for CPU time on the Brutus cluster, the
Santa Fe Institute for CPU time on the Scoville cluster, and the Centro
de Supercomputacion y Visualizacion de Madrid (CeSViMa) for access to
the Magerit cluster.  M.A.M.-D.~and H.B.~acknowledge financial support
from the Spanish MINECO Grant No.~FIS2012-33152, the Spanish MINECO
Grant No.~FIS2015-67411, and the CAM research consortium QUITEMAD+,
Grant No.~S2013/ICE-2801.  The research of M.A.M.-D. has been supported
in part by the U.S.~Army Research Office through Grant No.~W911N
F-14-1-0103.  H.G.K.~acknowledges support from the National Science
Foundation (Grant No.~DMR-1151387) and the Swiss National Science
Foundation (Grant No.~PP002-114713).  Part of the research of H.G.K.~is
based upon work supported in part by the Office of the Director of
National Intelligence (ODNI), Intelligence Advanced Research Projects
Activity (IARPA), via MIT Lincoln Laboratory Air Force Contract
No.~FA8721-05-C-0002. The views and conclusions contained herein are
those of the authors and should not be interpreted as necessarily
representing the official policies or endorsements, either expressed or
implied, of ODNI, IARPA, or the U.S.~Government.  The U.S.~Government is
authorized to reproduce and distribute reprints for governmental purpose
notwithstanding any copyright annotation thereon.

\bibliography{refs}

\end{document}